\documentclass[10pt,twocolumn]{wlscirep}
\usepackage{mystyle}
\graphicspath{{img/}}
\usepackage{siunitx}
\usepackage{mathtools}
\usepackage{siunitx}

\title{On the 96-well plate coverglass tilt and curvature suppression in 96-camera imaging system}

\author[1]{Antony C.~S.~Chan}

\affil[1]{Current address: San Diego, CA 92131.}

\keywords{96-well plate, optical metrology, Fourier ptychography, High-throughput imaging}

\begin{abstract}
The 96-eyes instrument is capable of computational extended depth of focus (eDOF) of up to $\pm 30\mu m$ in the phase channel, and conventional depth of field (DOF) of $\pm 5\mu m$ in the fluorescence channel.
However, it requires minimal plate-to-plate cover glass depth variation to function. Plate depths are measured using a third-party plate scanner (Opera Phenix) grouped by plate types (UV-Star, Cell-Star, and Eppendorf meniscus-free).
The two-dimensional (2D) depth dataset is aggregated through principal component analysis to obtain the top eight dominating 2D surface deformation modes.
More than 90\% of the variation can be explained by the plate's absolute depth and tilt (Pitch, Gradient-Y, and Gradient-X), followed by ($\approx 2\%$) the cover glass's curvature (Curve-Y and Curve-XY).
Plate-to-plate average depth and tilt variations are suppressed by a customized kinematic mount anchoring the plate's cover glass at the instrument's imaging plane.
The plate's average curvature is compensated by manually aligning all 96-eyes microscope objective lenses to track the plate's surface; an one-off calibration procedure aided by the backlash-free piezo-flexure z-stage.
Design validation is conducted \emph{in silico}, with the proof of concept experiment conducted on the 96-eyes with new mounting bracket retrofits.
\end{abstract}

\begin{document}
\def\um{\,\si{\micro\meter}}
\def\mm{\,\si{\milli\meter}}
\def\nm{\,\si{\nano\meter}}
\def\Hz{\,\si{\hertz}}

\maketitle
\thispagestyle{fancy}

\section{Multi-modality, parallel, and lateral motion free microscopy for 96-well plates}

\begin{figure*}
\centering
\includegraphics[width=0.8\textwidth]{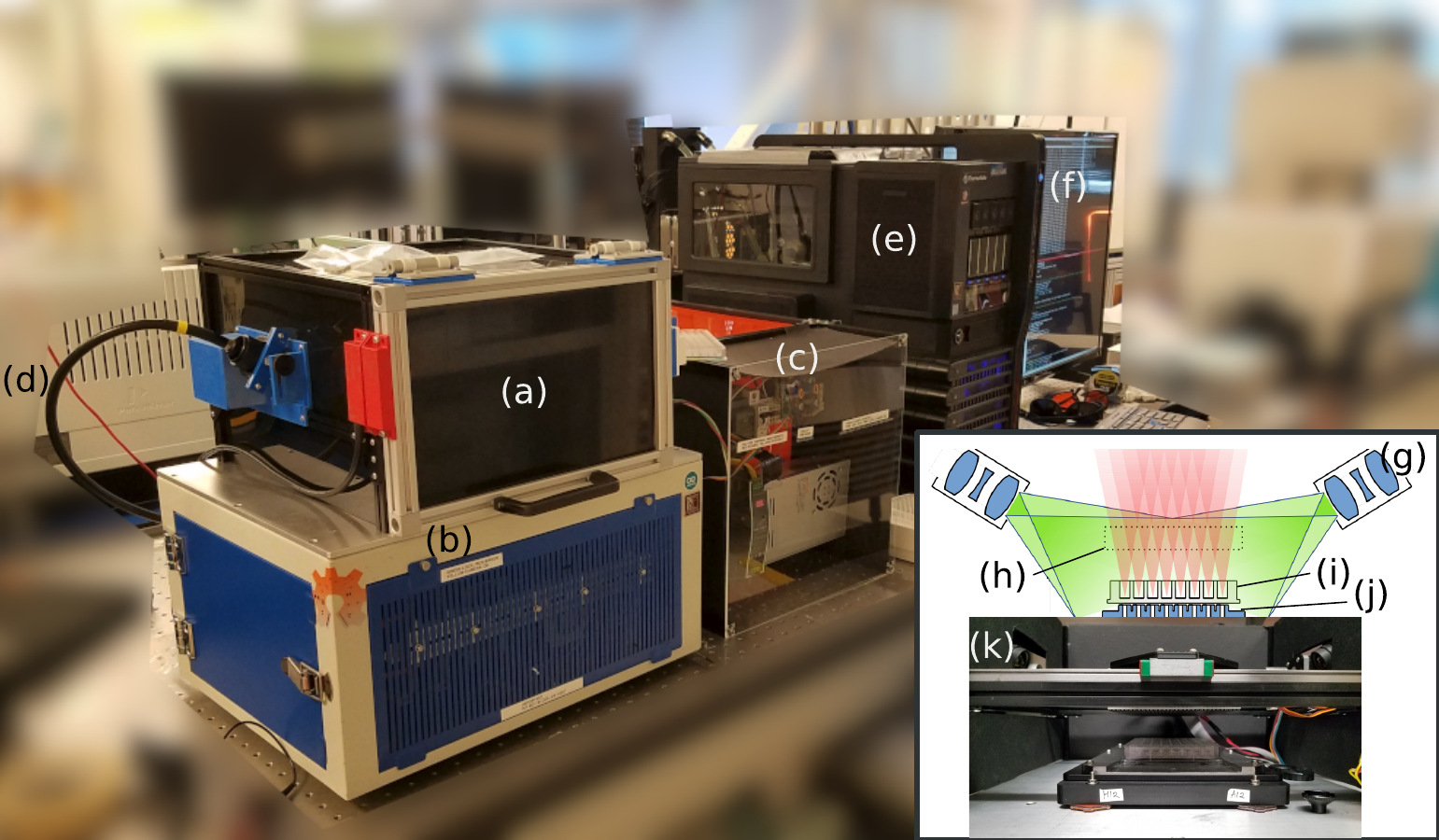}
\caption{\label{fig:setup}%
The 96-eyes intrument's latest iteration consists of
(a)~plate cradle and incubation module;
(b)~96-camera and power distribution module;
(c)~external illuminator as the dual-wavelength laser sources ($465\nm$ and $520\nm$),
transmitting excitation lights through (d)~a pair of homogenizing light pipes;
(e)~multi-GPU compute workstation; and
(f)~the 49-inch screen.
(Inset)~The front view of the plate cradle and the illustration.
(g)~Projection lens pair at the distal end of the homogenizing light pipes;
(h)~LED matrix illuminator;
(i)~96-well plate;
(j)~96-microscope objective array.
The new design also incorporates (k)~a customized piezo-flexure wide aperture stage for z-stack fluorescence imaging.
Computational phase imaging remains completely motion-free thanks to the Ptychographic extended depth of focus feature.}
\end{figure*}

The 96-eyes imaging system is a parallel motion-free microscopy system for high-throughput screening, capable of simultaneously imaging all wells on the 96-well cell culture plate. Since its first inception \cite{Chan2019}, the 96-eyes instrument underwent an upgrade to incorporate an additional fluorescence channel (excitations $= 465\nm,\ 520\nm$, emission $= 510\nm,\ 625\nm$), as well as the ability to conduct an one-off z-stack fluorescence image capture step per plate (Figs.~\ref{fig:setup}) with post-acquisition autofocusing.

Enabled with the new motion and illuminator hardware, we revisited the earlier claim of lateral-motion-free multimodal imaging with Ptychographic extended depth of focus. In particular, we wish to address the effect of plate-to-plate cover glass variations in the new setup without the need for motorized lateral (i.e., x,y axes) or goniometric (i.e., pitch and roll) position controls.

\begin{figure*}
\centering
\includegraphics[width=0.8\textwidth]{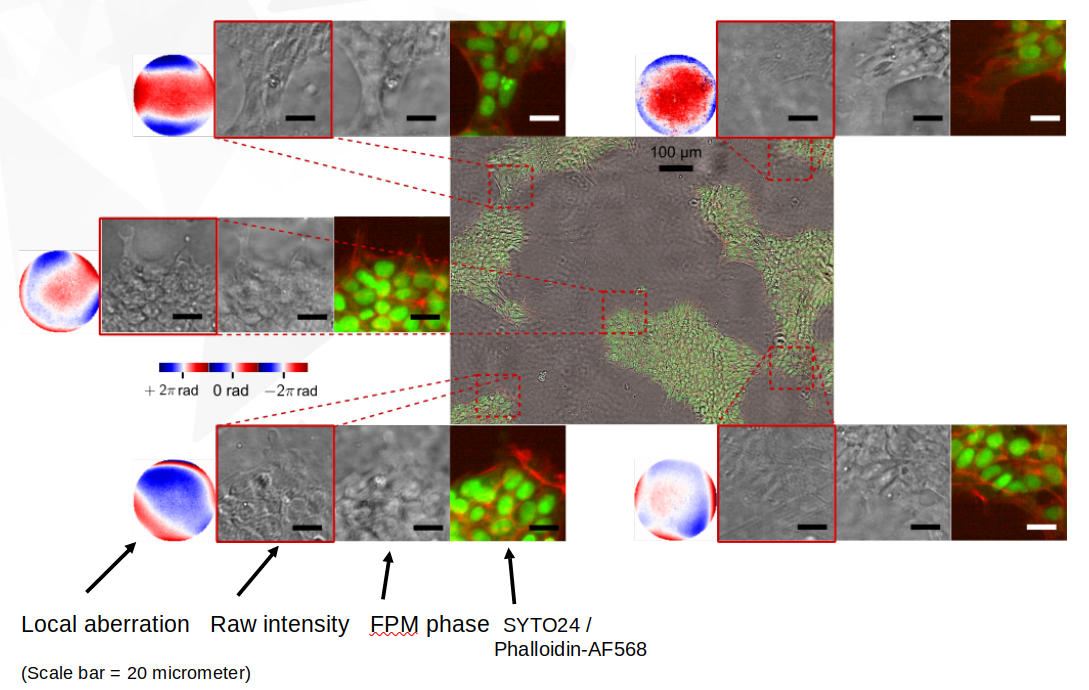}
\caption{\label{fig:dual-channel}%
Multi-modality imaging capabiltiy of individual imaging light paths for each well.
The 96 individual light paths acts as individual mini-microscopes, enabling parallel imaging with computational optical aberration suppression.}
\end{figure*}

\section{Prior studies on well depth variation characterization}

\begin{figure*}
\centering
\includegraphics[width=0.8\textwidth]{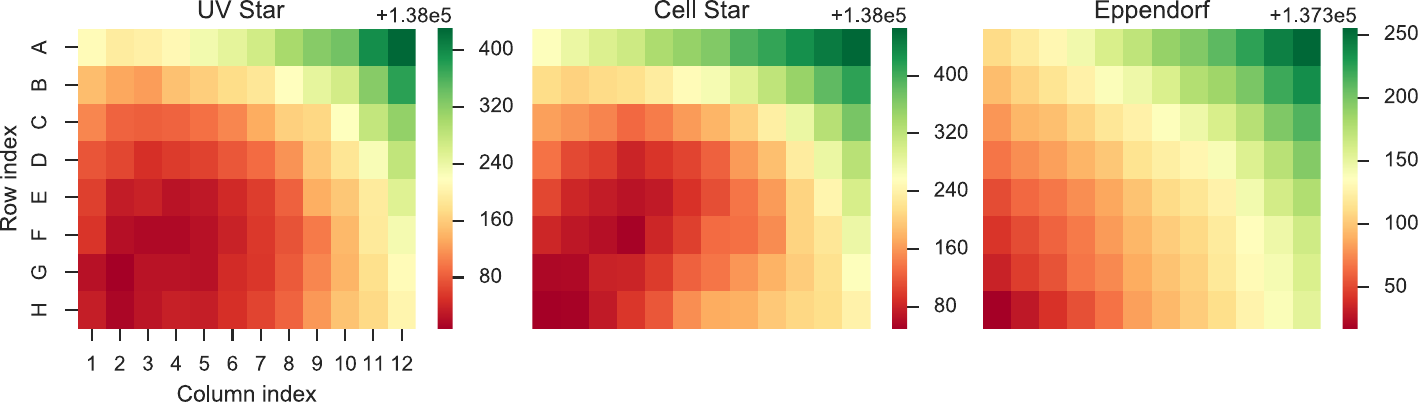}

\includegraphics[width=0.4\textwidth]{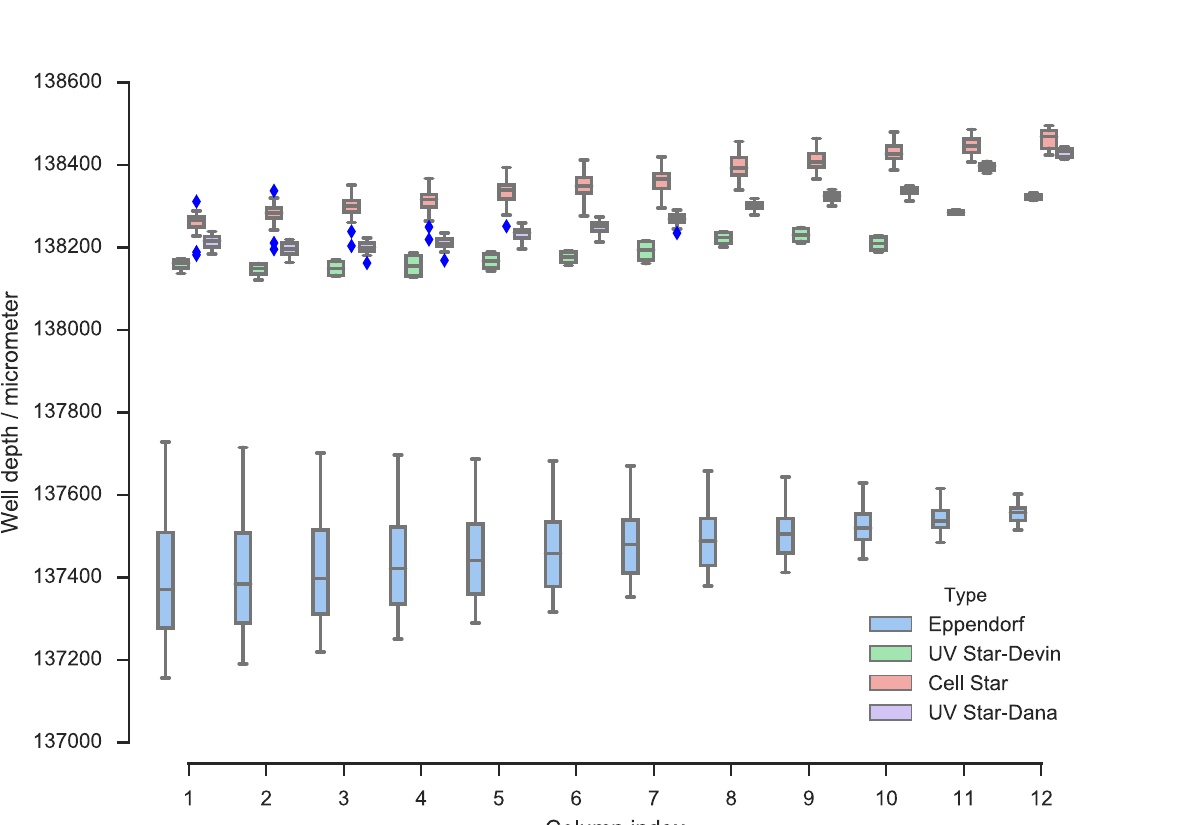} \includegraphics[width=0.4\textwidth]{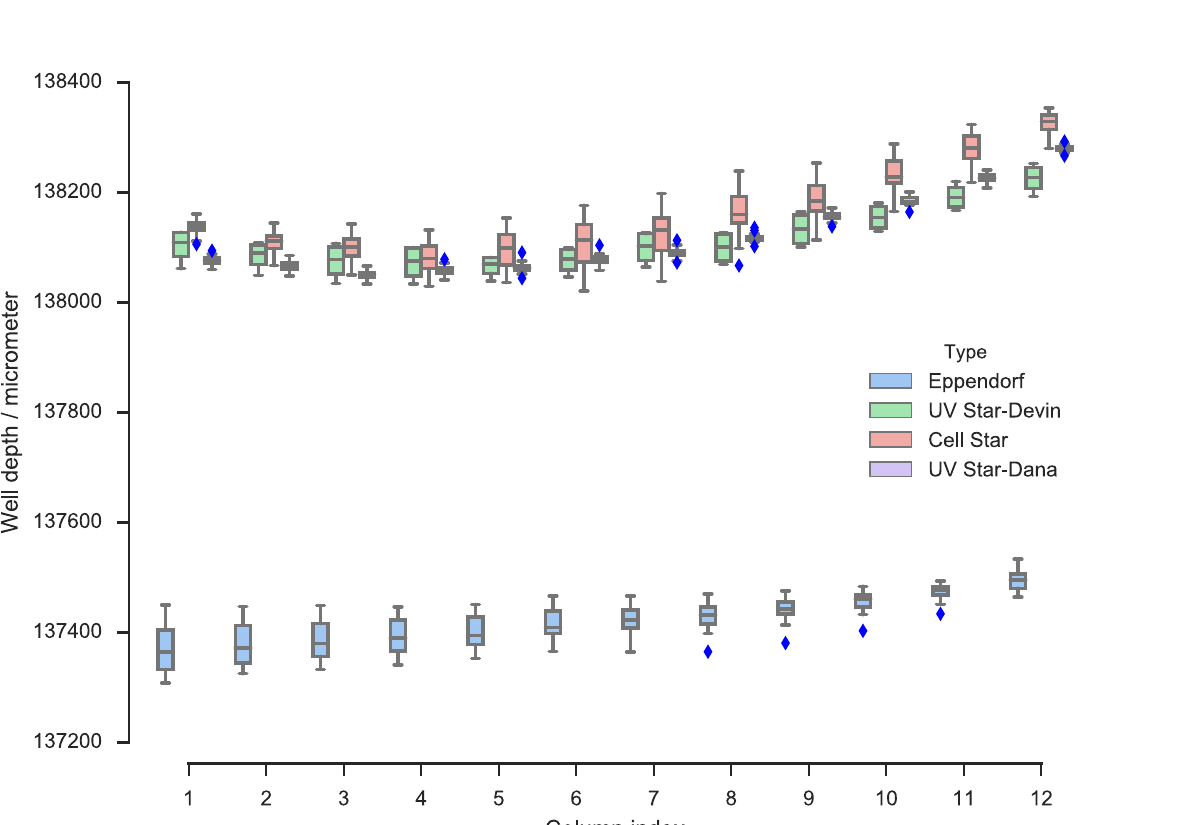}

\caption{\label{fig:average-height}%
(Top)~Average coverglass surface profiles.
(Bottom left)~Box plot of the depth distribution for Row A, and (Bottom-right) Row D.}
\end{figure*}

Previously, we characterized the depth variations by analyzing 16~plates (UV-Star plates) at room temperature using an Opera Phenix high-content imaging system equipped with laser-based autofocus. The UV-Star plates are known for their thicker chimney side walls and the optical-quality polymer bottom at $0.17\mm$ thickness. For fair comparisons, we also evaluated 16~generic culture plates with polystyrene bottoms (Cell Star plates), as well as 22~photochemistry plates having an approximate bottom thickness of $1\mm$ (Eppendorf plates). As shown in Fig.\ref{fig:average-height}, all plate types exhibit a general curvature of the well bottom with a depression of up to $500\um$. A strong tilt of the well bottom is also observed in the depth profile. This implies a gentle but significant plate tilt relative to the flange of the 96-well plate where the former sits. Previously in the study, we asserted that such \emph{per well} average tilting and curvature can be compensated with an one-off manual focusing step for individual microscope objectives.

\section{Plate tilt as the top contributing factor of depth variation}

We applied \emph{principal component analysis} (PCA) to determine the top contributing factors to height variations.
The dataset representing the measured plate depths (dimensions: plate count $\times$ row count $\times$ column count) is first arranged in lexicological order, having a two-dimensional (2D) matrix with dimensions of (plate count $\times\ 96$).
The matrix dimension is reduced to $8 \times 96$ with PCA by selecting the top eight principal components.
The values in the matrix are then rearranged back to $8 \times 8 \times 12$ to represent the stack of 2D plate depth profiles\cite{Turk1991}.

\begin{figure}
\includegraphics[width=\columnwidth]{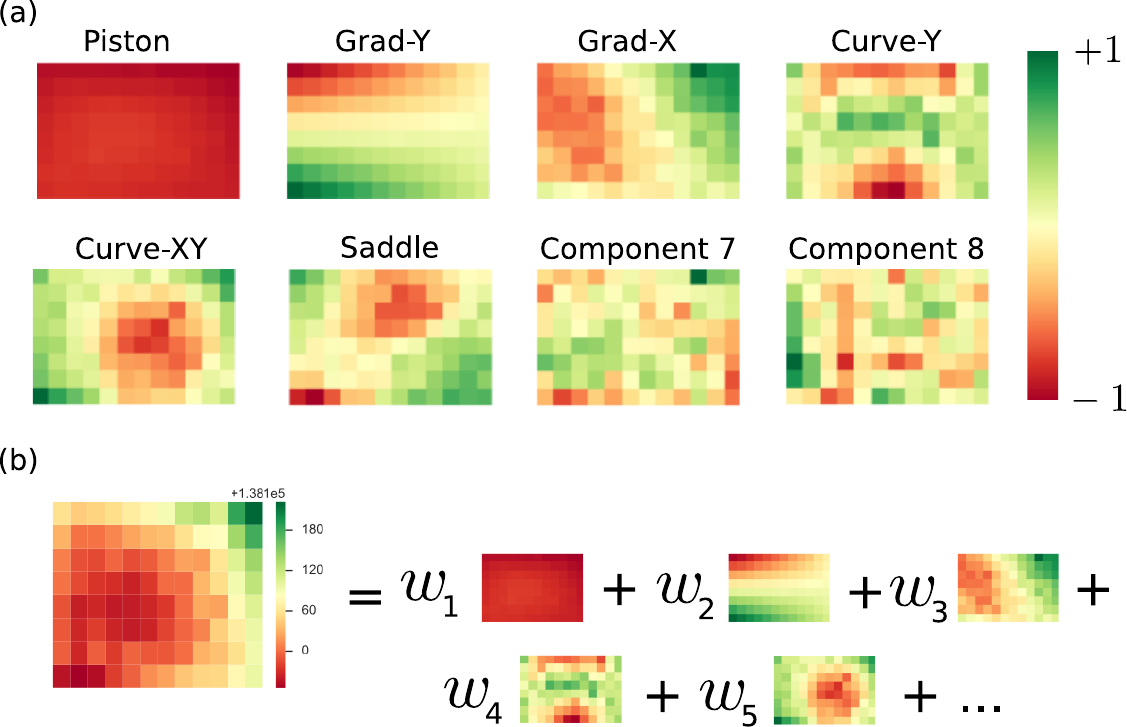}

\caption{\label{fig:principal-components}%
(a)~Top eight contributing factors of the 96-well plate depth variations.
For the sake of clarity, each components are named by the respective characteristic shape:
Piston, Gradient-Y, Gradient-X, Curve-X, Curve-XY, and Saddle.
The colorscales are normalized and non-dimensional.
(b)~Any particular well plate's coverglass surface profile can be approximated as the \emph{weighted sum} of the top five dominating components,
where the weights (unit: micrometers) indicates the relative importance.}
\end{figure}

The top eight most dominant plate depth surface ``modes'' are illustrated in Fig.~\ref{fig:principal-components}(a).
For the sake of clarity, the modes are named based on the common understanding of the characteristic shapes, namely Piston, Gradient-Y, Gradient-X, Curve-X, Curve-XY, and Saddle.
For our stack of 96-well plates, the corresponding depth profile can be represented as the sum of all profile modes multiplied by the weights (Fig.\ref{fig:principal-components}).
By definition, PCA solves for depth profile modes such that the magnitudes of the weights are in exponential descending order.

\begin{figure}
\includegraphics[width=\columnwidth]{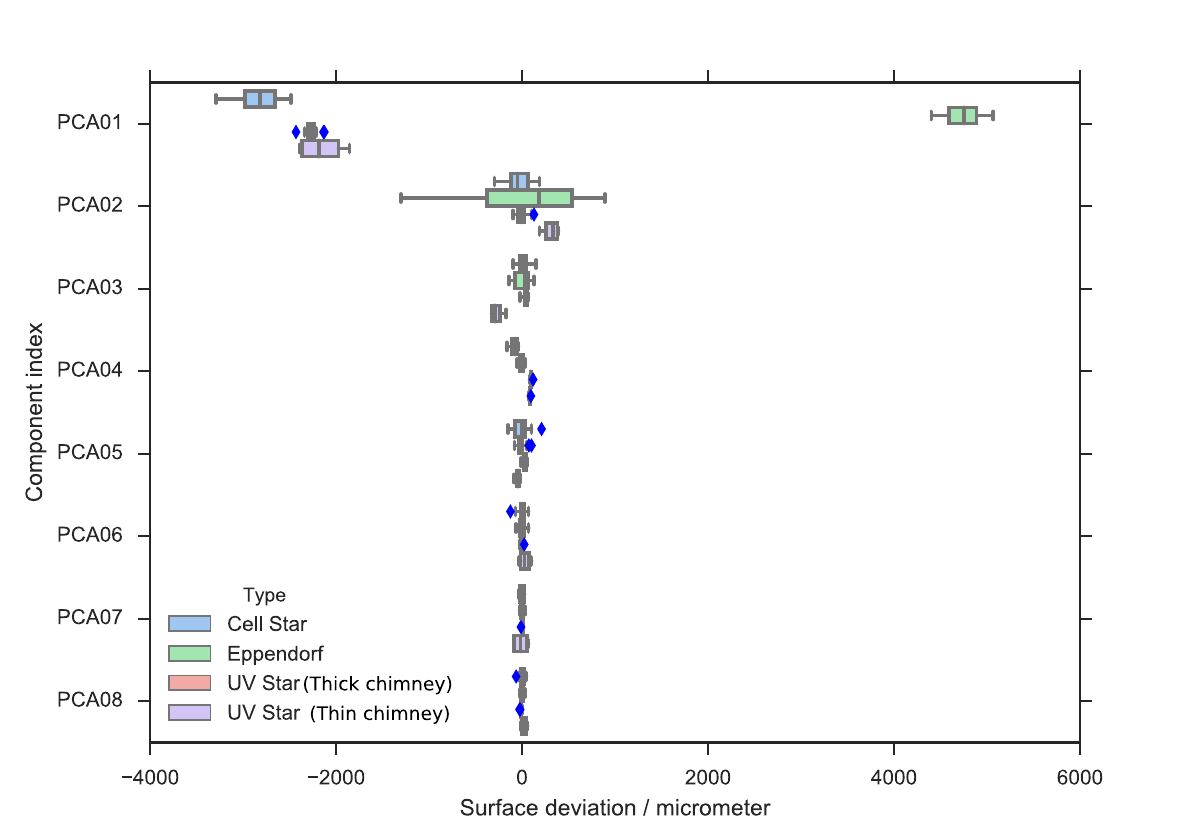}
\includegraphics[width=\columnwidth]{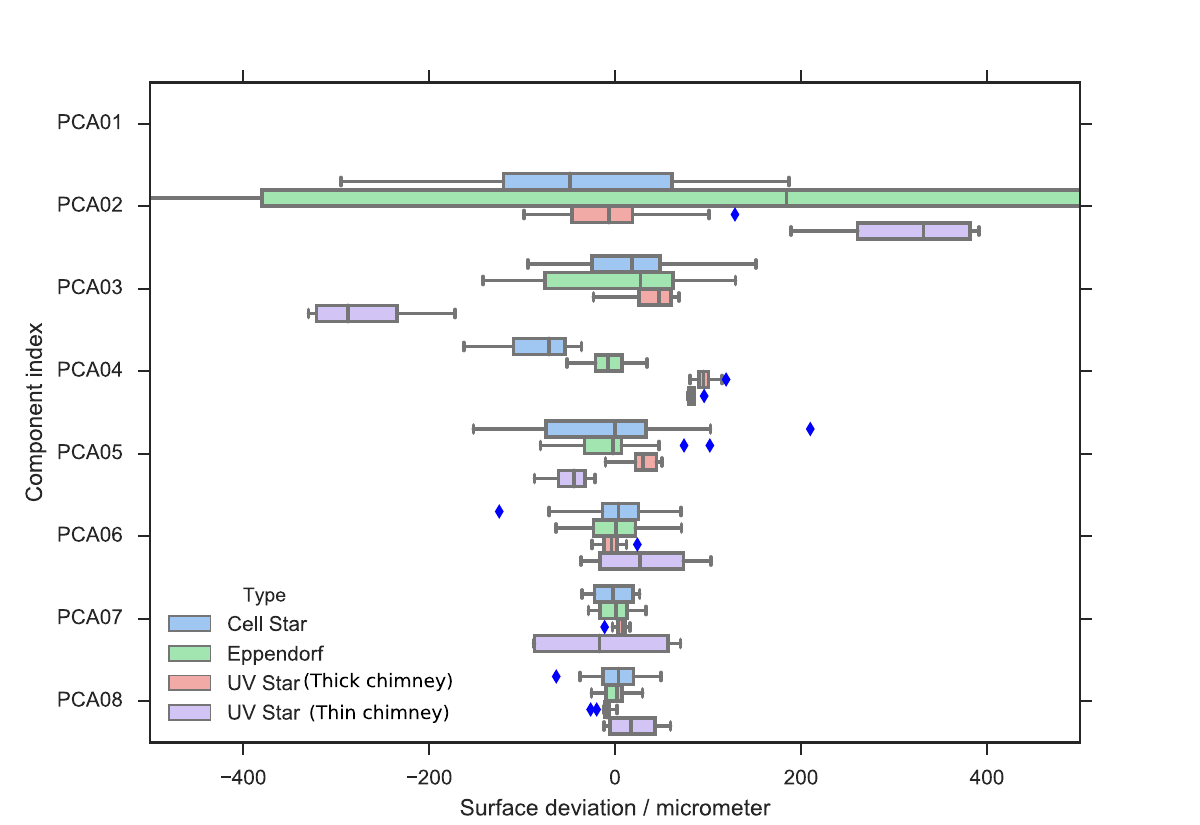}
\caption{\label{fig:PCA-weights}%
Statistical distributions of the PCA weights of indivdual 96-well plates, grouped by the plate type.
(a)~Weights of all eight principal components.
(b)~Close up view of the weight distributions for Grad-X, Grad-Y, Curve-Y, and Curve-XY.}
\end{figure}

The statistical distribution of PCA weights is illustrated as box plots in Figs.~\ref{fig:PCA-weights}(a--b), grouped by the plate types (UV-Star, Cell-Star, and Eppendorf).
Since various plate types have their unique \emph{flange-to-coverglass} distance defined in their specifications, we excluded the first principal component \emph{Piston} from our analysis.
From the box plots, it is revealed that the plate's tilt (Gradient-X and Gradient-Y) contributes to approximately 90\% of the plate-to-plate depth profile variations; another 2\% is contributed by the plate's curvature.
Comparing the tilt variations across plate types, the Polystyrene plates (Cell-Star) have the highest tilt variations, followed by Eppendorf plates and the UV-Star plates.

We also compared two sub-groups of UV-Star plates having wall thicknesses of $1.93\pm0.05\mm$ and $0.95\pm0.05\mm$.
The plates with thicker chimney walls exhibit around a 50\% reduction in the tilt and curvature variations [Fig.~\ref{fig:PCA-weights}(b)].

\section{Motion-free plate tilt suppression with kinematic mounts}

The 96-eyes project's vision is to combine the strengths of parallel camera hardware and computational extended depth of focus technology,
so we can eliminate the complex motion hardware as required by all conventional single-camera systems.
X-Y axes stages, especially those having fast settle time, high precision, and low position drift,
are known to be cost-prohibitive in comparison to the consumer-grade CMOS image sensors and plastic-molded microscope objectives.
Complex motion control algorithms are finely tuned to avoid sloshing liquids in the 96-well plates.
Therefore, plate tilt suppression with motorized goniometer controls is rejected early in the design roadmap---
with the exception of z-axis motorized control.
It is only until recently (June 2019), the backlash-free, wide aperture z-axis stage
($1\mu m$ resolution, $300\mu m$ travel range, $250\Hz$ open-loop response, ZSA300, DSM, Franklin, TN) were integrated into the instrument for dual-fluorescence z-stack imaging.

In order to suppress plate-to-plate tilt variation, we drew inspiration from the kinematic mounting mechanism,
where the mounting tool's tip and tilt are precisely determined on three alignment pins of the baseplate.
We started with simulation of the kinematic mount idea by placing the pins strategically at three chimney locations:
(i)~between well D1 and E1; (ii)~between well A12 and B12; and (iii)~between G12 and H12.
The mounting pins are located under the chimneys so that we do not sacrifice any wells for imaging.
We also took advantage of the 96-eyes instrument's small imaging ROIs by design; only the center field of view having diagonal distance of $\approx 1.5\mm$ is captured by the camera.

\begin{figure}[b]
\includegraphics[width=\columnwidth]{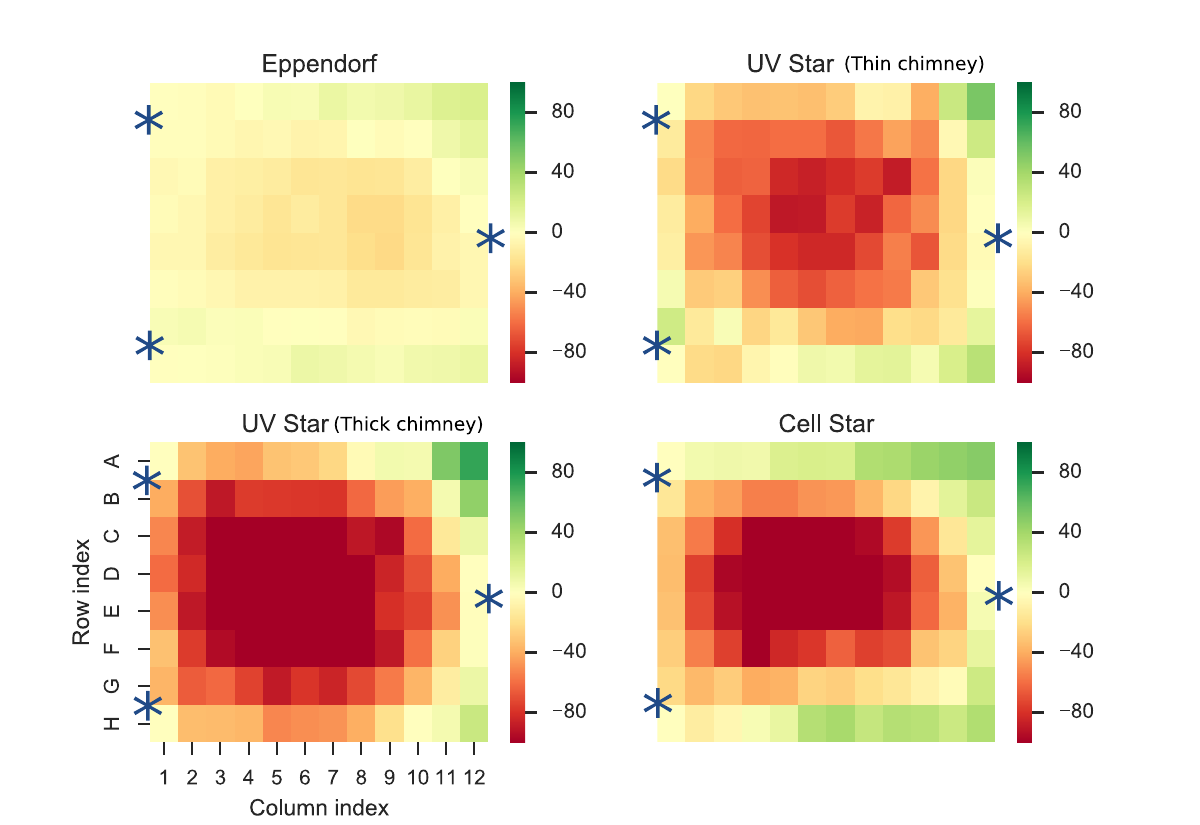}

\caption{\label{fig:sim-kinematic-mount}%
Simulating motion-free tilt suppression by mounting the plates's coverglass on the kinematic mounts.
The blue asterisks illustrates the positions of the mounting pins of the simulated kinematic mounts.
Since the pins are located underneath the chinmey walls, no wells need to be sacificied in the imaging step.}
\end{figure}

\begin{figure}
\includegraphics[width=\columnwidth]{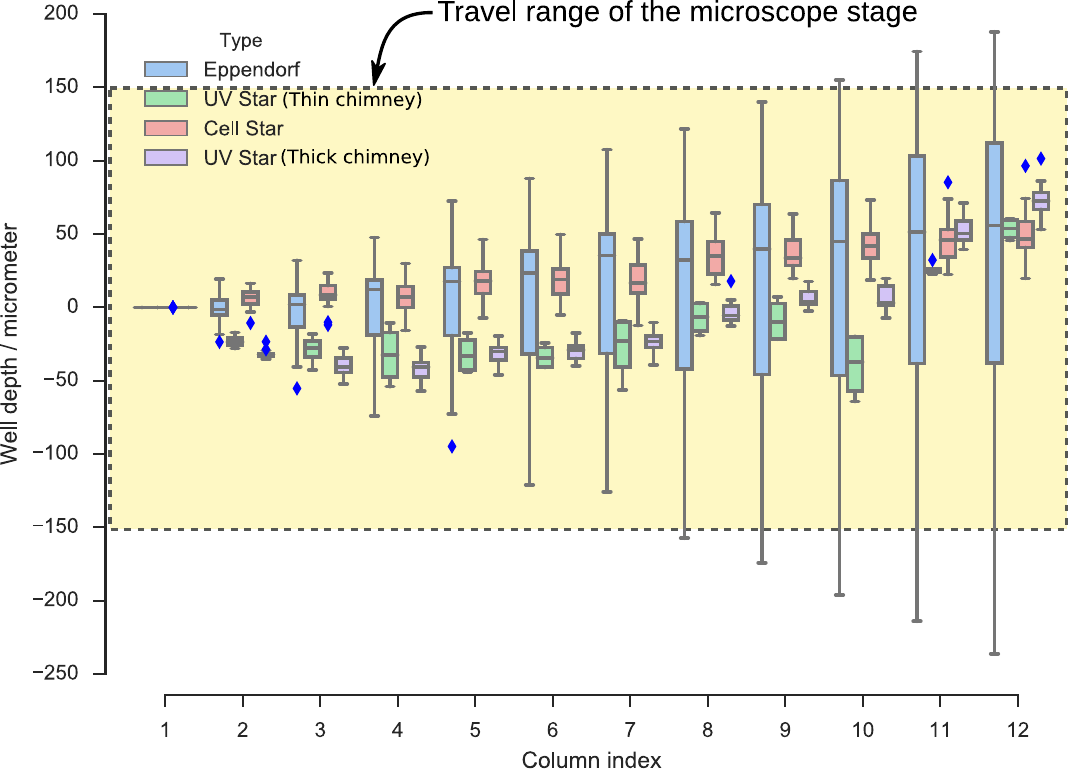}
\includegraphics[width=\columnwidth]{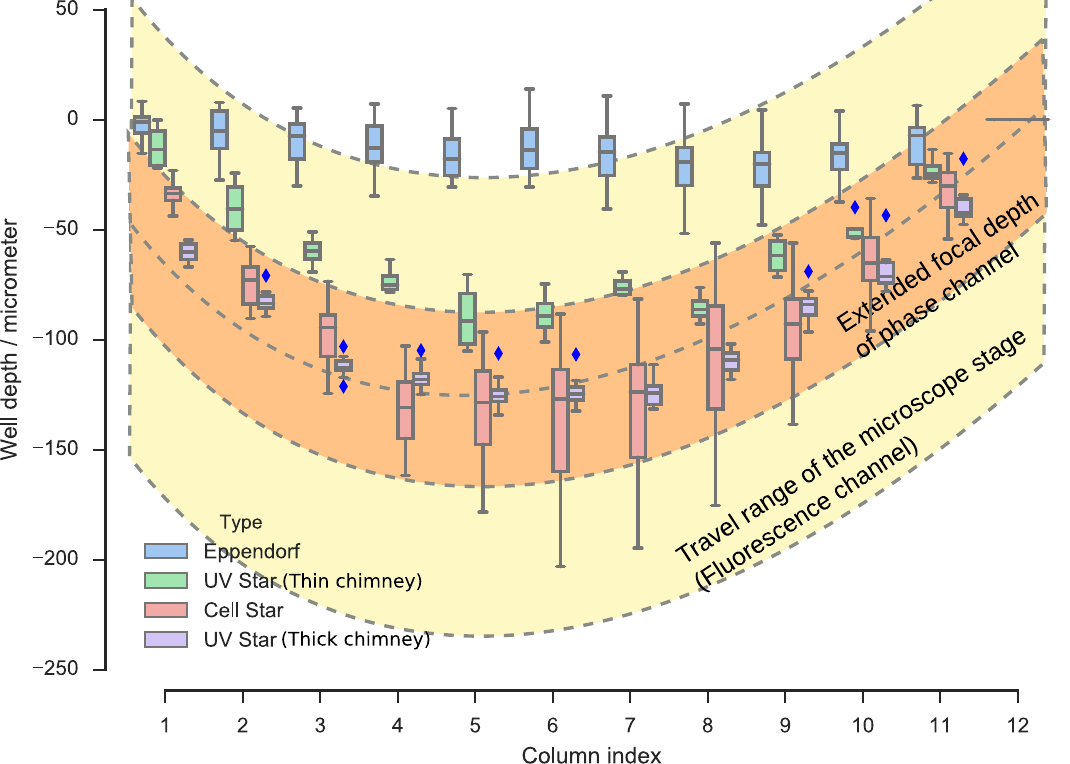}
\caption{\label{fig:sim-leveled-boxplot}%
Depth distribution of simulated motion-free tilt suppression.
(Middle)~Simulated depths for Row A; and (Bottom)~for Row D.}
\end{figure}

To simulate the effect of plate tilt suppression, we process the plate depth measurements by computing the \emph{global tilt gradient}
from the triangulated locations~(i) to~(iii), and then subtract the plate depths \emph{in silico} (Fig.~\ref{fig:sim-kinematic-mount}).
We also plotted the distributions of the compensated depths under the backdrop of the z-stage's maximum close-loop travel range ($300\mu m$),
as well as the instrument's extended depth of focus range in the computational ptychographic phase channel (Fig.~\ref{fig:sim-leveled-boxplot}).

\begin{figure}
\includegraphics[width=\columnwidth]{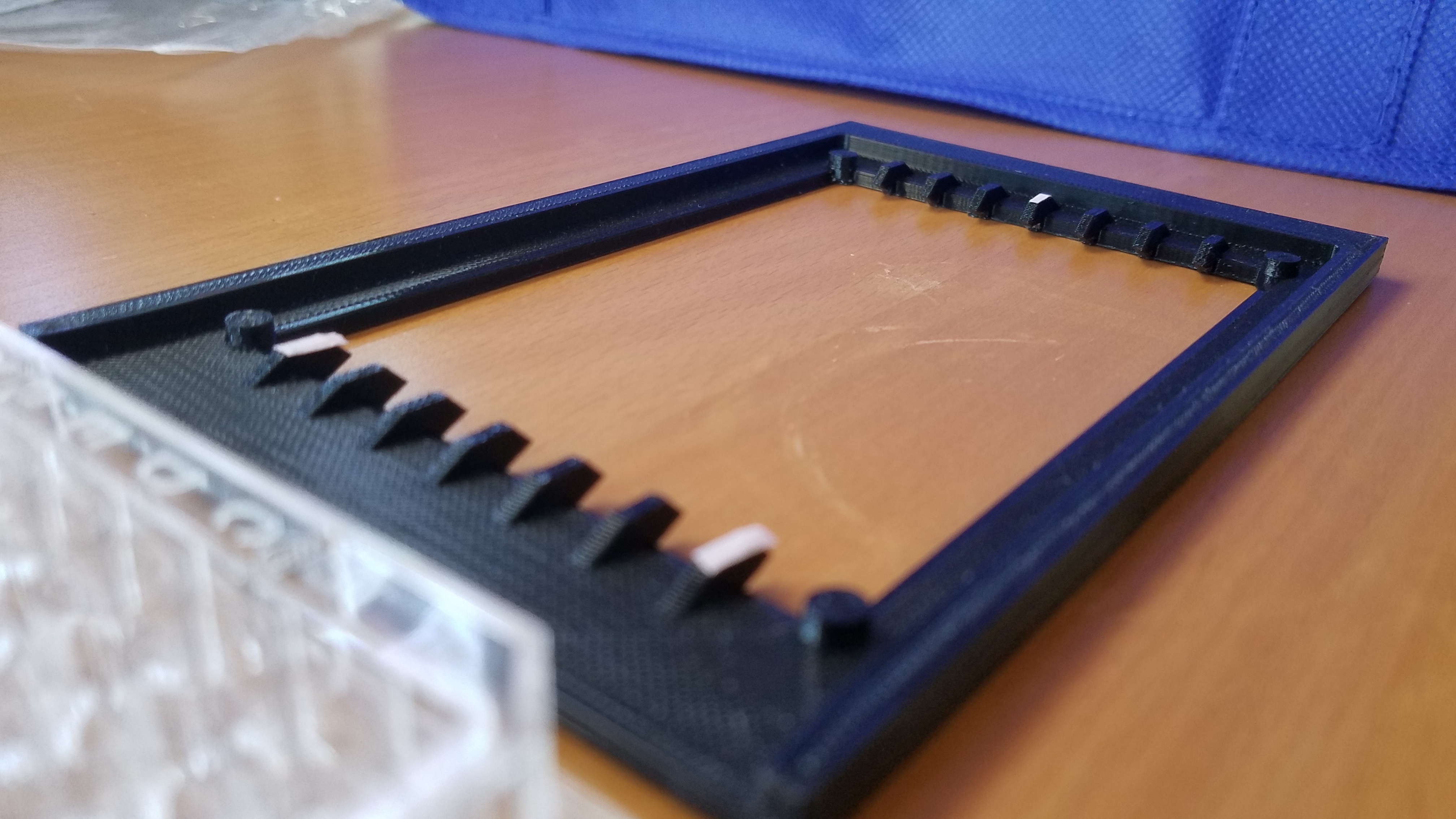}
 
\caption{\label{fig:bracket}%
Early prototypes of the kinematic mounting mechanisms for tilt suppression.
Later designs have the 3D printed alignment pins glued onto the machined polycarbonate sheet.
There are plans to replace 3D printed pins with adjustable M2 setscrews.}
\end{figure}

The kinematic mounting mechanism was eventually implemented as a machined z-stage bracket out of a polycarbonate sheet, which 3D printed alignment pins glued onto
the locations~(i) to~(iii). The pins are 3D printed and glued rather than snap-fit with a setscrew because of the now obsolete requirement front-loading of the 96-well plates; the plates were required to slide in from the front entrance, raised on the 3D printed ramps, and then dropped into position on top of the pins (Fig.~\ref{fig:bracket}).

We did not repeat the depth measurement experiment on the Opera Phenix plate depth scanner because unlike 96-eyes, the former does not permit custom bracket bolted onto the loading platform without voiding warranty.

\section{Plate curvature supression with one-off microscope lens distance calibration}

\begin{figure}
\includegraphics[width=\columnwidth]{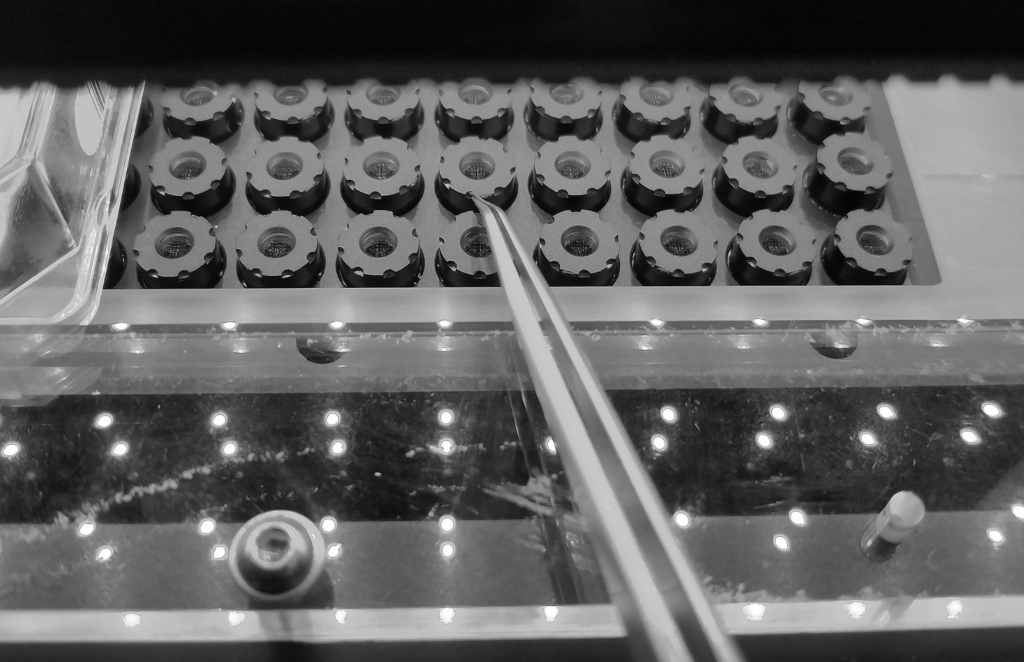}
 
\caption{\label{fig:lens-barrel}%
Manual, one-off microscope focuser actuation with lens thread.}
\end{figure}

\begin{figure*}
\includegraphics[width=\columnwidth]{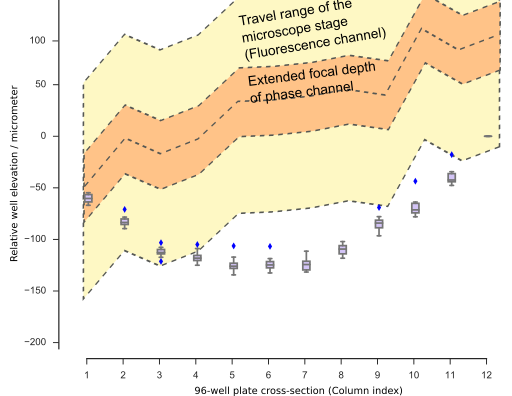}
\includegraphics[width=\columnwidth]{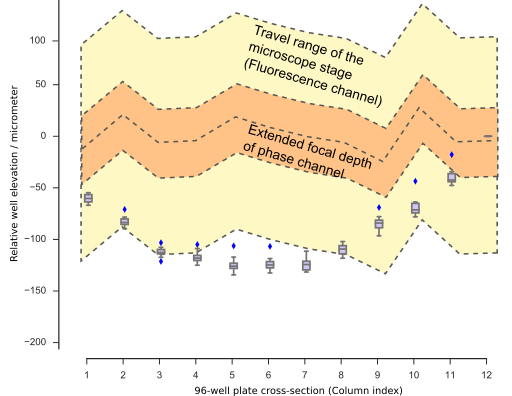}
\includegraphics[width=\columnwidth]{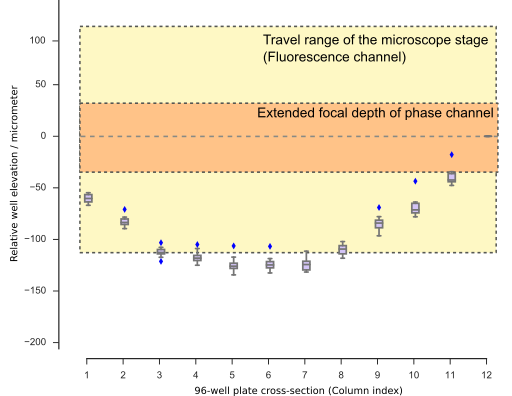}
\includegraphics[width=\columnwidth]{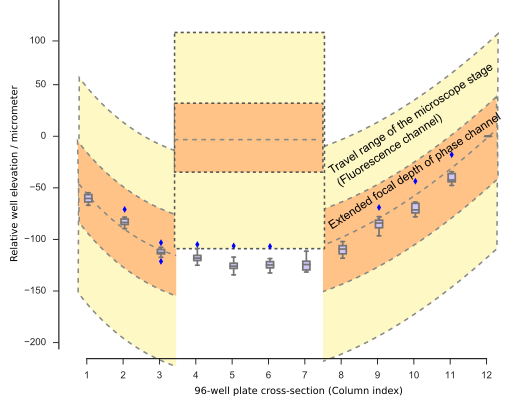}
\includegraphics[width=\columnwidth]{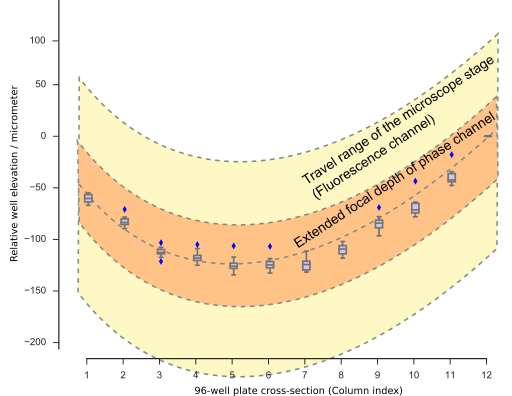}
 
\caption{\label{fig:lens-curvature-calibration}%
One-off plate curvature suppression workflow through manual lens position adjustment.
From left-to-right, then top-down:
Initial random focal planes of the 96-eyes instrument;
After installing the kinematic mount and leveling the z-axis stage;
After resetting the microscope lens vertical positions;
After manually suppressing focal planes variation through single z-stack phase image capture;
Expected post-calibration lens positions and the corresponding focal planes with respect to the plate's curvature.}
\end{figure*}

As the 96-well plate's cover glass is mounted on top of the custom kinematic bracket having 3 alignment pins,
the plate-to-plate tilt variation drops below the plate's average curvature.
The residual imaging defocus distance is now dominated by the plate's average curvature.
Here, we devised the manual calibration procedure to adjust individual lens barrels' parfocal distances, so as to track the 96-well plate's average curvature
(Fig.~\ref{fig:lens-curvature-calibration}).
The 96-eyes instrument, as the name suggests, is composed of 96 individual imaging light paths having independent control of lens vertical positions (Fig.~\ref{fig:lens-barrel}).
As of the time of writing, we do not find any commercial solution of motorized actuators having a form factor of $9\mm$ and the circular aperture of $6\mm$,
but we anticipate the motor actuation technology will be available in the market through miniaturization and wafer-level motor-optics integration.

The plate-to-plate's curvature variation is addressed by 96-eye's computational extended depth of focus.

\section{Summary}

The plate-to-plate depth variation is addressable with the pre-factory, one-off calibration procedure in the 96-eyes instrument.
Through the plate depth measurements, we identified the best plate type (i.e., having the lowest depth variations) for the 96-eyes instrument.
We also characterized the plate's 2D depth profile via principal component analysis, so we can identify the most dominant surface ``modes'', i.e., Pitch,
Gradients, Curvature.
We address the plate tilt variations with a motion-free kinematic mount having alignment pins touching the plate's cover glass under the chimneys.
We also suppressed plate-to-plate average curvature by an one-off microscope lens' parfocal distance adjustment.
Integrated with the backlash-free precision z-axis stage, we can capture stain-free cell cultures without any X-Y-tilt motions,
as well as dual fluorescence channel support via z-stack imaging to counteract spectral focal shift.

\end{document}